\documentclass{article}

\usepackage[utf8]{inputenc} % allow utf-8 input
\usepackage[T1]{fontenc}    % use 8-bit T1 fonts
\usepackage[colorlinks=true,linkcolor=blue, citecolor=blue, urlcolor=blue]{hyperref}       % hyperlinks
\usepackage{url}            % simple URL typesetting
\usepackage{booktabs}       % professional-quality tables
\usepackage{amsfonts}       % blackboard math symbols
\usepackage{nicefrac}       % compact symbols for 1/2, etc.
\usepackage{microtype}      % microtypography
\usepackage{graphicx}
\usepackage{comment}
\usepackage{lmodern,textcomp}
\usepackage{subcaption}
\usepackage{amsmath}
\usepackage{amsbsy}
\usepackage{longtable}
\usepackage{adjustbox}
\usepackage{lscape}
\usepackage{verbatim}
\usepackage{float}
\usepackage{booktabs}
\usepackage{dcolumn}
\usepackage{comment}
\usepackage{multirow}
\usepackage{amsthm,amsmath}
\usepackage{tabularx}
\usepackage{authblk}
\usepackage[section]{placeins}
\usepackage{natbib}
\setcitestyle{authoryear,open={(},close={)}}
\graphicspath{{images/}}

\title{Simulating Tertiary Educational Decision Dynamics: An Agent-Based Model for the Netherlands}

\author[1]{Jean-Paul Daemen}
\author[2,*]{Silvia Leoni}
\affil[1]{\small School of Business and Economics, Maastricht University, Maastricht, The Netherlands}
\affil[2]{\small Department of Economics, Statistics, and Business, Universitas Mercatorum, Rome, Italy}
\affil[*]{\small Email: \href{mailto:silvia.leoni@unimercatorum.it}{silvia.leoni@unimercatorum.it}}

\date{}

\begin{document}

\maketitle

\begin{abstract}
This paper employs agent-based modelling to explore the factors driving the high rate of tertiary education completion in the Netherlands. We examine the interplay of economic motivations, such as expected wages and financial constraints, alongside sociological and psychological influences, including peer effects, student disposition, personality, and geographic accessibility. Through simulations, we analyse the sustainability of these trends and evaluate the impact of educational policies, such as student grants and loans, on enrollment and borrowing behaviour among students from different socioeconomic backgrounds, further considering implications for the Dutch labour market.
\end{abstract}

{\bf Keywords:} higher education, agent-based modeling, decision-making.

\section{Introduction}
The Netherlands has exhibited a remarkable increase in tertiary education attainment among young adults (aged 25–34), rising from 44.3\% to 56.4\% in the past decade—a growth rate surpassing that of top-performing EU countries such as Norway \citep{eurostat2023attainment}. While the Dutch trajectory suggests convergence with high-attainment nations, similar starting positions a few years back in countries like Belgium and France have not yielded comparable progress. This divergence raises critical questions: What drives the Netherlands’ accelerated growth in higher education participation, and are these trends sustainable?

Existing literature offers varied perspectives on the determinants of tertiary education enrollment. Survey-based approaches, such as those by \cite{sojkin2012determinants}, highlight familial influence and personal aspirations as key factors in Poland. Economic models, notably \cite{becker1992go} human capital theory, emphasize wage premiums, whereas \cite{khan2022choice} incorporates demographic disparities, including gender and socioeconomic background, though without elucidating underlying mechanisms. Agent-based modeling (ABM) provides a dynamic alternative, capturing social interactions in educational decision-making \citep{manzo2013educational,leoni2022agent}. However, these studies predominantly focus on low- or moderate-enrollment contexts (e.g., Italy), leaving high-growth systems like the Netherlands underexplored. Notably, Dutch-specific research remains scarce, with only limited attention to geographic accessibility \citep{sa2006does}.

This study addresses these gaps by employing an ABM framework to analyze the social and economic drivers of the Netherlands’ exceptional enrollment growth. We extend the literature in three ways: (1) identifying unique determinants in a high-attainment setting, (2) evaluating the sustainability of current trends, and (3) assessing labor market implications of rising educational supply. Our findings offer policymakers insights into reinforcing equitable access while mitigating potential oversupply risks.

We develop a highly data-driven model able to reproduce the observable figures for enrolling rates in the Netherlands. Our model shows that these high figures can realistically be reproduced by taking into account the complex interaction of economic, sociological and psychological factors which all contribute to educational decision-making, albeit the economic driver appears to be the most relevant. Results indicate a significant difference in the student loan amount between first-generation highly educated students and those with a highly educated family background, with the latter relying more on the credit system. This result is in line with observed data and literature showing how students from educated families can usually rely on better financial stability which allows for easier access to borrowing.
The model is then employed to conduct a scenario analysis that allows to test the efficacy of national policies directed at increased access to higher education. 

Agent-based modeling is uniquely suited to investigate the dynamics of educational attainment. Firstly, it allows for a greater level of complexity than standard economic models. ABM can include features and variables which are more challenging to include in other models. Secondly, the interaction component allows influences from other agents and the environment. Both of these reasons improve the model’s ability to simulate reality. Lastly, adapting the restrictions offers the possibility to test regulatory changes without the need for additional data.

Recent applications of agent-based modeling (ABM) have provided valuable insights into educational choices and systemic outcomes. In the domain of career selection, \cite{mozahem2022social} developed an ABM framework incorporating four social cognitive theories to examine gender disparities in STEM fields. The simulation demonstrated how subtle initial biases in self-efficacy and social expectations compound through social interactions, resulting in significant underrepresentation of women in technical disciplines - a finding that challenges conventional linear regression analyses. Complementing this work, \cite{allen2010simple} employed ABM to analyze declining STEM enrollment in U.S. universities, revealing a nonlinear relationship between perceived costs and enrollment decisions. Their most striking finding indicated that marginal reductions in perceived barriers (as little as 10-15\%) could produce more than twofold increases in STEM major selection, highlighting the importance of threshold effects in educational decision-making.

The flexibility of ABM has proven particularly valuable in modeling college selection processes. \cite{reardon2014agent} constructed an agent-based framework to disentangle the complex interplay between family socioeconomic status and college application behaviors. Their results not only confirmed the primary role of financial resources but also quantified how peer networks and information asymmetry moderate this relationship. \cite{diaz2021agent} further advanced this line of inquiry by simulating the transition from neighborhood-based to choice-based school allocation systems. Their model captured how information diffusion about school performance metrics reshaped enrollment patterns, leading to increased concentration of students in higher-achievement institutions - an emergent phenomenon that could not have been predicted through traditional analytical methods.

ABM has proven its strength also in examining the dynamics and evolution of education markets. \cite{maroulis2014modeling} developed a sophisticated model to examine school choice mechanisms, uncovering a paradoxical temporal mismatch between parental preferences and educational outcomes. Their model explores how household-level decisions, specifically the emphasis on school achievement versus proximity, impact district-wide educational outcomes. Surprisingly, the simulations reveal a counter-intuitive finding: a greater emphasis on achievement by parents does not consistently lead to higher average district achievement. This arises partly due to the timing of new, higher-quality schools entering the system, which can be negatively affected by parents' immediate pursuit of existing high-achieving schools. This counterintuitive finding underscores ABM's ability to capture complex temporal dynamics in educational systems.

Foundational work by \cite{manzo2013educational} established the critical role of social networks in educational transitions, demonstrating through computational experiments that individual ability and cost-benefit calculations alone cannot explain observed enrollment patterns. \cite{leoni2022agent} subsequently refined this framework, focusing specifically on tertiary education decisions and introducing policy-testing capabilities. The current study extends this line of research by adapting the ABM approach to the Dutch context, where unique cultural factors and exceptionally high attainment rates may alter the relative importance of various determinants. Furthermore, we incorporate labor market feedback mechanisms, enabling analysis of how educational policy changes propagate through both the education system and employment sectors.

The paper proceeds as follows: Section \ref{sec:factors} reviews the literature investigating the variables affecting educational decision-making which are integrated in the model illustrated in Section \ref{sec:model}. Section \ref{sec:simulation} guides through the initialization and dynamics of the simulated model. Results are presented in Section \ref{sec:results} and complemented by a discussion of the Dutch labour market trends in Section \ref{sec:labor}. A scenario analysis and a sensitivity analysis are performed in Sections \ref{sec:scenario} and \ref{sec:sensitivity}. Section \ref{sec:conclusion} concludes and presents potential extensions of the present analysis.

\section{Factors influencing educational decision-making}\label{sec:factors}
A multitude of factors can influence an individual's decision to enroll in university. These factors often interact in complex ways, making it challenging to disentangle their individual effects. This section aims to identify the most relevant and appropriate variables for inclusion in the agent-based model.

\subsection{Wage premium}\label{subsec2.1}

Following the human capital theory \citep{becker1964human}, first among other factors contributing to educational decision-making is the wage premium. Empirical evidence from the Dutch labor market shows significant heterogeneity across age, groups, gender, and social backgrounds, and in particular, occupation, with large variation between educated and practically skilled jobs \citep{schneck2021trends, allen2020educational}. Recent research shows how construction occupations reach wage premium levels of several educated jobs \citep{EIB2022}. 

\subsection{Parental income}\label{subsec2.2}

Education serves as a powerful yet imperfect mechanism for social mobility. While higher education attainment strongly correlates with increased lifetime earnings \citep{arifin2017role}, socioeconomic factors such as parental income may significantly constrain access to tertiary education itself \citep{jackson2020century}. This is particularly pronounced in systems with high tuition fees and quality variation between institutions, such as the United States, where \cite{jackson2020century}'s (2020) cohort analysis revealed that economic expansions indeed significantly reduce enrollment inequalities. 

The mechanisms perpetuating this cycle are well-documented. Affluent families leverage their resources through superior college preparation via private schooling \citep{bastedo2011running}, access to tutoring and application guidance \citep{haveman2006role}, and better financial resources \citep{goldrick2016reducing}.

European systems with lower income inequality and tuition fees show similar patterns. French data reveals a 5.8\% enrollment increase per 10\% income redistribution \citep{bonneau2022unequal}, with parental transfers and information asymmetry playing key roles comparable to US findings. Even in the Nordic countries characterized by tuition-free universities and generous grant systems, significant income-based disparities persist. Danish data shows top-income children maintain 50\% higher enrollment rates \citep{thomsen2015maintaining}, challenging assumptions about universal welfare states.
Considering the Dutch context, it is important to mention that the government switched from a basic grant to a social loan system in 2015. At the start of the academic year 2023-2024, the system was reversed into a basic grant regulation due to concerns regarding debt accumulation \citep{van2022encouraging}. Previous to the implementation, questions were raised about whether the impact of socio-economic background would increase. To investigate the parental income effect, the Central Agency of Statistics tracked student achievement for a cohort born in 1989 \citep{kazemierouderlijk}. This study shows that for similar secondary educational achievements, children from wealthier parents have a greater possibility of attending tertiary education. The difference between the tales of the distribution amounted to a difference close to 50 \%. Partially based on the results of the study, the ministry installed a parental income-based grant system to prevent the gap from widening. 

It should be noted that the study simply measures the correlation between the two variables and does not control for coinciding factors. In research conducted by the Council of Education, scholars demonstrated significant relations with confounding factors such as ability, demographics and migration background \citep{Onderwijsraad2011}. 

\subsection{Student disposition}\label{subsec2.3}
Conventional wisdom suggests natural intelligence should strongly predict college enrollment and subsequent success, yet empirical evidence reveals a more complex relationship. \cite{richardson2015does} conducted a comprehensive meta-analysis of IQ-performance studies, finding that previously reported correlations were often inflated by methodological artifacts. When accounting for measurement errors and controlling for psychosocial factors like self-esteem and self-efficacy, the relationship between IQ and job performance proved substantially weaker than traditionally assumed.

These findings align with more recent specialized studies. \cite{iqbal2021relationship} examined medical students in Lahore and found no significant correlation between IQ scores and academic performance, despite observed gender differences in IQ testing. Similarly, analysis of the Wisconsin longitudinal study revealed that while IQ showed some association with later earnings, this relationship was notably mediated by family background, high school environment, and particularly by academic effort as measured by class rank.

The limited predictive power of pure cognitive measures becomes especially apparent when comparing assessment methods. \cite{galla2019high} demonstrated that high school grades consistently outperform standardized admission tests in predicting on-time college graduation. This suggests that non-cognitive factors - including planning ability, self-regulation, and consistent effort - captured through sustained academic performance provide more meaningful predictors of educational success than intelligence measures alone.

Together, these studies indicate that while cognitive ability represents one component of academic potential, its predictive power is substantially moderated by psychosocial factors and learning behaviors. 

\subsection{Peers influence}\label{subsec2.4}

Peer influence in educational decision-making is complex and extends beyond simple imitation. Mimetic interactions, as described by \cite{manzo2013educational} , highlight how individuals’ likelihood of enrollment increases with the observed choices of peers, positioning social influence alongside academic ability and perceived returns as central to educational stratification.

Empirical studies underline the nuanced nature of these effects. \cite{hallinan1990students} found that reciprocal and cross-gender friendships exert stronger influence on college aspirations, while racial disparities persisted despite high aspirations among African American students. \cite{holland2011power} further showed that peer norms can both encourage and undermine college preparation efforts within minority communities.

Experimental evidence \citep{andersen2022unnoticed} suggests that peer influence often operates unconsciously, challenging self-report-based research. Contextual variation is also significant: German studies report divergent peer effects depending on tracking systems and ability grouping, while gender composition has shown mixed results across educational levels \citep{cabezas2010gender, OOSTERBEEK201451}.

Students from disadvantaged backgrounds are especially susceptible to peer effects due to limited access to institutional support and familial college knowledge. Peer networks can significantly increase college attendance rates among these groups \citep{sokatch2006peer}, although immigrant students face additional barriers due to social exclusion \citep{plenty2017social, rosenqvist2018two}.

Finally, social comparison processes add another layer of complexity. Exposure to high-achieving peers may either boost or diminish aspirations depending on individual performance levels and gender dynamics \citep{jonsson2008choice, modena2021does}. Negative peer experiences, such as college rejection, can also shape others’ enrollment decisions, particularly within similar demographic groups \citep{pistolesi2022enrolling}.

Overall, peer influence emerges as a multifaceted and context-sensitive driver of educational choices, mediated by social networks, institutional structures, and individual characteristics. 

\subsection{Personality}\label{subsec2.5}

A growing body of research examines how personality traits, particularly the Big Five dimensions, influence educational pathways and academic success. 
Longitudinal evidence from \cite{coenen2021personality} suggests that openness to experience positively predicts STEM specialization, with extraversion, agreeableness, and neuroticism showing negative associations. The effect of openness is especially pronounced among female students, indicating potential relevance for understanding gendered enrollment patterns in higher education.

While direct evidence linking personality traits to college enrollment is limited, \cite{loche2017psychological} found that conscientiousness and agreeableness were positively associated with enrollment among U.S. military veterans. However, the specificity of this sample limits broader generalizability to student populations.

Stronger evidence exists for personality-academic success relationships. \cite{trapmann2007meta}'s meta-analysis found conscientiousness to be a consistent predictor of academic achievement, while neuroticism correlated negatively with academic satisfaction. Openness showed no clear relationship with grades, though data on its impact on satisfaction were inconclusive.

In contrast, \cite{chen2022relationships} identified positive associations between openness—particularly to diversity and intellectual challenge—and first-year college outcomes, including improved peer interaction, course engagement, and second-year retention. These findings suggest openness may play a role in academic integration and persistence.

Overall, personality traits contribute in differentiated ways to educational trajectories. Conscientiousness reliably predicts academic performance, while openness appears more relevant for educational decision-making and adjustment.

\subsection{Geography}\label{subsec2.6}

The Netherlands exhibits significant geographic disparities in educational attainment, with metropolitan areas outperforming rural regions \citep{CLO2023}. According to Kooiman (2018), this gap arises primarily from post-secondary migration rather than initial regional differences in educational potential. While high school students nationwide exhibit comparable promise, graduates disproportionately relocate to the Randstad in pursuit of better wage growth and career opportunities.

University proximity also plays a significant role in enrollment decisions, albeit at a local scale. \cite{sa2006does} show that improved access, particularly to research universities, reduces secondary school dropout rates. These localized effects often go undetected in regional analyses. International evidence from Germany confirms distance becomes a critical factor beyond 12.5 km due to increased transactional costs \citep{spiess2010does}, though cross-country comparisons require caution given varying national contexts.

These findings underscore two geographic dimensions influencing educational outcomes: long-term migratory patterns affecting human capital distribution, and localized access barriers shaping initial enrollment. Both have critical implications for regional policy and educational infrastructure planning.

\section{Model Design}\label{sec:model}

Drawing from \cite{leoni2022agent}, the model is populated by two generational groups: the \textit{juniors}, those facing the choice to enrol in university, and the \textit{seniors}, already engaged in the labour market as educated or practically skilled. In other words, students versus working residents.
As the society examined is relatively large, the model must have a low network density where only a few potential links are created \citep{hamill2009social}. Furthermore, The model is instructed to facilitate high clustering, so agents should connect with those geographically proximate. 
This is ensured by using the structure of social circles \citep{hamill2009social}. Each node can solely create links with other nodes within a given radius. This is referred to as the 'social reach' of an individual. Consequently, individuals with a greater social reach will have a larger social network. Concerning seniors, the model employs a fat-tailed distribution of connectivity to demonstrate the differences in social activeness among individuals. The drawback of using such a distribution is that individuals may perceive others as their social connections, but this perception might not be mutual. Since students need reciprocal links with other students within their reach to determine peer influence, their social reach is set to a constant value.

\subsection{Budget Constraint} \label{sec:budget_constraint}

When students evaluate their decision to enrol, they account for multiple factors. Within the model, students first construct a budget on the basis of their financial resources and compare it with educational costs. The educational costs depend on whether a prospective student aspires to live independently, which will be the case for 53 per cent of students \citep{nji2023kameers}. If so, Table \ref{table3} shows that this will increase expenses by an average of 695 euros for groceries and rent. In total, parental and outliving students are projected to face costs of 749 and 1444 euros respectively \citep{Nibud2021}.

Conversely, college students will receive income from several sources. Firstly, from the academic year 2023-2024 onwards students will receive a basic grant which amounts to 121.33 euros for parental living students and 302.39 euros for outliving ones \citep{duo2023}. Secondly, depending on parental income, students will be entitled to a supplementary grant with a maximum of 457,60 euros. The full amount will be granted for a household income up to 36592.92 and subsequently linearly decreases until it reaches zero around 80.000 euros \citep{duo2023}. Considering that the average Dutch household has a gross income of 75.200 euros, and follows a right-tailed distribution, a significant number of students qualify for at least a part of the additional grant \citep{CBS2022InkomenHuishoudens}. 

Furthermore, parents financially support their children's education as observable in Table \ref{table2}. The table is constructed by combining Central Bureau of Statistics (CBS) data on gross household income and German research that examined the spending behaviour on children's higher education in Germany \citep{CBS2022InkomenHuishoudens, kornrich2013investing}. Most noteworthy is the observation that lower-earning parents spend a larger share of their income on children's education. Resultingly, only high-earners contribute significantly more to education in absolute terms.

In addition, the Netherlands reports the highest share of students who are employed during their studies for all EU member states \citep{Eurostat2022workstudents}. Specifically, 72 per cent of Dutch students receive an income from work earning on average 508 euros monthly. In comparison, Danish and German students are in second and third place with 49 and 42 per cent respectively. The student income data follows a log-normal distribution since most students earn marginally above minimum wage, while a small part earns significantly more \citep{kleiber2003statistical}. 

Lastly, if the resources mentioned above will not suffice, students qualify to apply for a student loan with a monthly maximum of 1054.17 \citep{DUO2023lenen}. In theory, prospective students who cannot cover expenses with the mentioned resources will not be able to attend college. According to \cite{Studiefinanciering2020}, this virtually never occurs in practice. The main purpose of constructing the budget within the model is to simulate loan behaviour, which serves as an input in the expected consumption function. Therefore, budget constraint is as follows: 

\begin{equation}
    X_{i,t} - E_t > 0, \label{eq:1}
\end{equation}

Where $X_{i,t}$ represents the financial funds for individual $i$ in year $t$. $E_t$ depicts the costs of studying for students in year $t$. 

If the result is negative, the student will automatically enrol in the job market as practically skilled. The second hard condition is related to the overall result of the high-school exams. If these are insufficient, a high-school student will retry to succeed in the next period. In the following subsections, the variables of the preference equation are established. This equation forms the input for calculating the likelihood of entering college. 

\subsection{Consumption}

Firstly, the preference equation includes an equation for expected consumption to incorporate the impact of enjoying a wage premium after college completion. It is hypothesised that an individual's decision to attend university is influenced by its potential profitability \citep{becker1964human}. Therefore, the equation is calculated as a ratio of educated over practically skilled expected consumption. The real expected consumption for educated workers at time $t + 1$ takes the form:

\begin{equation}
    C^{e}_{ie,t+1} = Y^{e}_{ie,t+1} - L_{it} \label{eq:2}
\end{equation}

the corresponding equation for practically skilled:

\begin{equation}
    C^{e}_{ip,t+1} = Y^{e}_{ip,t+1} \label{eq:3}
\end{equation}

where $Y^{e}_{ie,t+1}$ and $Y^{e}_{ip,t+1}$ represents the average expected income for educated and practically skilled workers, as in \cite{leoni2022agent}. $L$ depicts the additional costs of repaying potential accumulated debt, including interest. 

Family background forms an important factor in determining wage expectations \citep{MENON2012805}. In the equation below, parental income receives more weight than others' wages, as students have grown up observing their family's income and regard it as a benchmark. This effect is likely stronger if students wish to pursue the same discipline as their mother or father \citep{brunello2001wage}. Therefore, a standard neighbour link receives $\nu = 1$, whereas the parents' weight equals $\nu = 1 + r$ \citep{leoni2022agent}. The supplementary weight $r$ is drawn from an even distribution $U(0,1)$, implying the parental effect has maximal twice the weight. 

Additionally, the family's place of residence and social network greatly determine the social network students rely on when forming their wage expectations \citep{MENON2012805}. The model reflects this by locating students near their parents, ensuring a significant overlap in their social reach. 

Derived from the above, the following expected wage function is defined for skilled labour: 

\begin{equation}
     Y^{e}_{ie,t+1} = \frac{\sum^{n\upsilon}_{i=1}\nu Y_{nWor,lh,t}}{n\upsilon} \label{eq:4}
\end{equation}

Where  $nWor$ stands for neighbours active in the skilled labour market at time $t$. 

In the model, educated labour constitutes 36 per cent of the total labour market, with no seniors being unemployed. These working neighbours are randomly assigned an occupation that falls into either the low- or high-earning category for educated workers (eqq. \ref{eq:4} and \ref{eq:5}). This assignment reflects the reality that higher education does not guarantee above-average earnings \citep{Beroepen2023}. These categories are established using Table \ref{table1}, highlighting considerable wage disparities between occupations. 

A similar function is used to estimate the expected income from practical labour. The function also differentiates between low- and high-income occupations as shown in Table \ref{table1}. Furthermore, to reflect the high wages earned by self-employed construction workers, the model assigns 3,6 per cent of practically skilled seniors an average gross monthly wage of 7500 euros at set-up \citep{EIB2022}. While construction jobs offer earnings similar to and above several educated occupations, services and manufacturing workers seem to fall behind. The complete equation is as follows:

\begin{equation}
     Y^{e}_{ip,t+1} = \frac{\sum^{n\upsilon}_{i=1}\nu Y_{nWor,lh,t}}{n\upsilon} \label{eq:5}
\end{equation}

Eventually, agents will compare the expected consumption function out of practical and educated work:

\begin{equation}
   ln(\frac{C^{e}_{ie,t+1}}{C^{e}_{ip,t+1}}) = ln(\frac{ Y^{e}_{ie,t+1} - L }{ Y^{e}_{ip,t+1}}) \label{eq:6}
\end{equation}

\subsection{Student's Disposition \& External Influences}

Besides economic considerations, prospective students will also review whether they have the necessary abilities to succeed in college and achieve worthwhile returns. Capturing such considerations in an equation is challenging, as the literature review displayed that not only intellect but also related abilities such as discipline and stress resistance should be taken into account \citep{IQearnings}. Resultingly, the equation below does not measure IQ but includes high school grades as a proxy for student's disposition \citep{galla2019high}. It should be evident that those extremely (un)gifted with study-related dispositions are (un)extremely likely to enrol in university. Therefore, the equation has the following form:

\begin{equation}
     SD_{it} = \lambda^\kappa_{it} \label{eq:7}
\end{equation}

Where SD stands for students' disposition, $\lambda$ ranges between 0 and 1 depending on one's final grade and $\kappa$ is assigned the value 1.8. As in \cite{leoni2022agent}, this ensures increasing returns to scale for those with above-average grades. The relatively high value for $\kappa$ is validated by the low intrinsic motivation of Dutch students with around average grades. In the Netherlands, this is also referred to as 'de zesjescultuur', and refers to students whose primary interest is to pass with minimal effort, with no desire to understand the materials studied. On average, Dutch students spend 33 hours on their study per week \citep{monitor2022}. This is significantly less than the European standard and falls far behind the highest-ranked countries' average, which exceeds 40 hours per week. 

Additionally, a student's preference to enrol in higher education is significantly influenced by peers \citep{andersen2022unnoticed, holland2011power, roth2017interpersonal}. Noteworthy, outperforming (under-performing) peers are solely influential when the difference with their performance is not too large \citep{jonsson2008choice, modena2021does, pistolesi2022enrolling}. The equation below accounts for peers' influence by dividing class average to an individual's GPA, using the resulting ratio as a multiplying factor of the student's disposition. To consider dissuasion effects, ratios that fall within the lowest and upper 10 per cent are inverted.  Together, this results in the equation below, where PI stands for peers' influence.:

\footnotesize
\begin{equation}
     \text{PI} = \begin{cases} 
       \frac{GPA_{Class}}{{GPA}_{Self}} & \text{if } \frac{GPA_{Class}}{{GPA}_{Self}} \leq 90th \; \text{percentile or if } \frac{GPA_{Class}}{{GPA}_{Self}} \geq   \text{10th percentile} \\
       1 / \frac{GPA_{Class}}{{GPA}_{Self}} &  \text{if } \frac{GPA_{Class}}{{GPA}_{Self}} \geq 90th \; \text{percentile or if } \frac{GPA_{Class}}{{GPA}_{Self}} \leq   \text{10th percentile}
   \end{cases} \label{eq:8}
\end{equation} 
\normalsize

Furthermore, the literature review discussed the effect of personality on enrolment choice. Regarding the Big Five Traits, openness to experience seems most related to college experience as it is positively related to retention rates and engagement levels \citep{ chen2022relationships}. The effects hold equally for male and female students and values for openness seem randomly distributed among the population. Accordingly, we define the equation below:

\begin{equation}
     PER_{it} =  O_{it} \label{eq:9}
\end{equation}

Where $PER_{it}$ stands for personality and $O_{it}$, the value for openness, varies between (0,1) following a normal distribution independent of any other factors. 

Lastly, to incorporate the significant effect of distance on enrolment in higher education \citep{sa2006does}, the model includes a variable that captures whether a prospective student is centrally located or not:

\begin{equation}
     CEN_{it} = \frac{avg.distance_{it}}{max.distance} \label{eq:10}
\end{equation}

Where $CEN_{it}$ resembles centrality and $avg.distance$ reflects the sum of the distances to all available universities over the number of universities present. Additionally, $max.distance$ is the maximum distance possible that a prospective student potentially would need to travel. 

Concluding, under the condition that the budget constraint is met and the final exams are passed, the utility equation represents the preference of a Dutch high school student to enter tertiary education: 

\begin{equation}
     P_{it} = \omega_{1}* ln(\frac{C^{e}_{ie,t+1}}{C^{e}_{ip,t+1}}) + \omega_{2}* ((PI_{it} * SD_{it}) +  PER_{it} +  CEN_{it}) \label{eq:11}
\end{equation}

The function is divided into two parts: one represents the economic motivation, while the second block resembles the intangible and social factors included. The initial weight of the economic component is set to 0.75, implying a weight of 0.25 for the social part. On average, the relative impact of intrinsic and extrinsic motivation demonstrates a complex relationship with no obvious dominant factor for European countries \citep{motivationinex2009}. However, \cite{drobnivc2010good} use survey data to conclude that Nordic European workers emphasise job security and payment more than residents from other European countries. Resultingly, the economic part receives a higher weight but the sensitivity analysis in Section \ref{sec:sensitivity} is used to verify whether altering weights significantly impacts the final results.  

Lastly, the preference function accounts exogenously for drop-out rates among Dutch (applied) university students, where a drop-out is defined as someone who leaves a study program and does not complete any study within nine years of enrolment. Concluding, the preference function determines the students utility from registering in higher education and completing a university degree. In line with \cite{manzo2013educational}, the probability to enroll and complete a higher education degree is monotonically non-linear increasing with the level of preference and concave shaped:   

\begin{equation}
 Pr_{it}(complete) =  \frac{exp(P_{it})}{1+ exp(P_{it})} \label{eq:12}
\end{equation}

\section{Model Simulation}\label{sec:simulation}

\subsection{Initialization}

The model is simulated in NetLogo \citep{wilensky1999netlogo}. The first step within the environment is to set up the model's society, which implies the creation of 3000 seniors and 11 universities. These senior agents are randomly assigned a gender and an age between 25 and 34 years old. Furthermore, with a 36 per cent probability their education is set to graduated, whereas the remaining seniors are (semi)-vocationally educated \citep{educationGPS2023}. As the model design specifies, these agents are randomly assigned wages from either low- or high-earning occupations.

Additionally, each agent is assigned a social-reach value drawn from a normal distribution determining its social connectivity. In reality, the distribution of individuals' social networks is significantly fat-tailed \citep{hamill2009social}. Therefore, 5 per cent of the agents are assigned an outlier value, while 1 per cent receives an extremely high social reach value. 

The location of each agent within the world is directed by educational background. This ensures that the model reflects educational homophily among social relationships \citep{thomas2019sources}. The NetLogo world is segregated into a left and right half, where practically (high) skilled agents mainly occupy the left (right) side. A senior is placed among similarly educated with probability ($\rho_{i}$) as shown by the following equation \citep{spiess2010does}: 

\begin{equation}
    \begin{cases} 
      \rho_{i} = {1} \; \text{if} \; x_{i} < \rho_{s}   \\
       \rho_{i} = \frac{1}{2} \; \text{otherwise }  
    \end{cases}
\end{equation}

Where $x_{i}$ is drawn from a standard uniform distribution and $\rho_{s} \in [0,1]$ indicates the level of segregation, with the degree of segregation increasing when $\rho_{s}$ approaches 1. If $x_{i} > \rho_{s}$, then location is determined randomly. For clarity, the above equation can be rewritten as follows:\footnote{ \begin{equation*} 
      \rho_{i} = 1 * \text{prob}(x <  \rho_{i}) + \frac{1}{2} * \text{prob}(x > \rho_{i})
\end{equation*} 
      where $x_{i}$ is drawn from a standard uniform distribution such that $prob(x <  \rho_{i}) = \rho_{s}$ and $prob(x > \rho_{i}) = 1 - \rho_{s}$ providing the equation above \citep[see][]{leoni2022agent}.
 }

\begin{equation}
      \rho_{i} = \frac{1}{2} + \frac{1}{2} \rho_{s} \label{eq:13}
\end{equation}

Finally, Dutch research on the strength of educational homophily remains inconclusive.  \cite{van2019sociocultural} utilised educational background statistics to predict the probability of relocating to a different neighbourhood. The study found a significant negative relationship between educational homophily among neighbours and the probability of moving elsewhere. However, research specifically addressing the impact of education on social connections is lacking. Therefore, $\rho_{s}$ is assigned a value of 0.5 and its impact will be evaluated in the sensitivity analysis Section \ref{sec:sensitivity}.

Table \ref{tab:1} provides an overview of the model's inputs at setup.

\footnotesize
\begin{table}[ht]
\centering
\caption{Variables’ values used in the model set-up with data source and year of reference in parentheses}
\label{tab:1}
\begin{tabularx}{\textwidth}{lX}
\toprule
\textbf{Variable} & \textbf{Inputs and calibration} \\
\midrule
N. senior agents & 3000 \\
Social reach (students) & 4.5 \\
Index of segregation $p_s$ & 0.5 \\
Number of universities $p_s$ & 11 \\
Initial Proportion skilled/unskilled & 36\% \citep{CBS2022Hoogopgeleiden} \\
Endowment & See Table \ref{table2} \\
Cost of education home & 749€ per month \citep{Nibud2021}  \\
Cost of education outliving & 1444€ per month \citep{Nibud2021}  \\
Ability & Secondary school grade \citep{CITO2023} \\
Income distribution & Educated high-earning  

$\sim \text{Mean€, sd€}(5246.5, 1445.76)$ \\
 & Educated low-earning 
 
 $\sim \text{Mean€, sd€}(3665.71, 211.11)$ \\
 & Uneducated high-earning 
 
 $\sim \text{Mean€, sd€}(3059.43, 343.62)$ \\
 & Uneducated low-earning 
 
 $\sim \text{Mean€, sd€}(2514.14, 184.98)$ \\
 & Uneducated constructors 
 
 $\sim \text{Mean€, sd€}(7350, 634.29)$ \\
\bottomrule
\end{tabularx}
\end{table}

\normalsize

\subsection{Model Dynamics}

The creation of the senior agents occurs during the null period of the simulation (t=0). Afterwards, the model is run for 100 periods of time, where each period represents a year. At the start of each period, seniors will \textit{hatch} students at the pre-defined birth rate of 0.05. Every student will be positioned close to their parent such that most of their social network up until adulthood stems from their family background. Contrary to seniors, the social reach of a student is fixed at 4.5 to ensure reciprocity among students.\footnote{The number 4.5 relates to the spatial units within the world of NetLogo. It indicates that students can connect with other agents up to 4.5 patches away from their position. In perspective, the world is fixed to a size of 20 by 20.} Else, one student would identify another as a classmate but not vice versa, which would conflict when calculating peers' influence. Comparatively, a senior has an average social reach of 5.5, highlighting that students are still developing their social lives. 

Furthermore, students, who are now 17 years old and in their final year of high school, must decide between pursuing higher education or entering working life. Within the model, the first evaluation determines if a student fulfils the budget condition and obtains a sufficient final grade as described in Section \ref{sec:budget_constraint}. Subsequently, students evaluate expected income from scientifically and practically skilled labour, using equation (\ref{eq:4}) and (\ref{eq:5}). These are employed for the respective consumption functions (\ref{eq:2}) and (\ref{eq:3}), which in turn complete the inputs for the expected consumption premium (\ref{eq:6}). Simultaneously, a student's disposition (\ref{eq:7}) is established, peers' influence is computed (\ref{eq:8}), the degree of openness (\ref{eq:9}) and centrality (\ref{eq:10}) are set. Finally, their preference to enrol is established.

The prospective students who decide to enter the labour force are turned into seniors and the highest obtained educational level will remain unchanged at high school. This implies that their income will be drawn from the corresponding income distribution and these agents are relocated such that the social network is independent of their parent. Contrarily, those who enrol in higher education take 5 time steps to complete their studies before transitioning into seniors, whose education is defined as graduates with earnings defined accordingly. Similarly, these graduates are relocated in correspondence with the segregation level defined at set-up.

To prevent overpopulation within the model and slow down the simulation process, senior agents aged above 45 are removed from the population. Additionally, random seniors are withdrawn if the population exceeds 3500 individuals. It is important to emphasise that the population's dynamics are not intended to replicate the characteristics of the real Dutch population. The population is purely created to allow an interplay between seniors and students and students between themselves.

The model's behaviour and the robustness of its results are evaluated by executing a Monte Carlo experiment and simulating the model 100 times. Within NetLogo, a software tool named BehaviorSpace exists, which enables the creator to test the model for all imaginable combinations of parameter values. Afterwards, it provides a dataset with detailed results.

Table \ref{tab:2} shows a step-by-step guide of the simulation process.

\begin{table}[H]
\centering
\caption{{The simulation algorithm}}
\label{tab:2}
\footnotesize
\begin{adjustbox}{max width=\textwidth}
\begin{tabular}{lp{0.8\textwidth}}
\toprule
\multicolumn{2}{l}{\textbf{Step-by-step manual}} \\
\midrule
1. \textbf{Initialization} & \\
\hspace{1em} (a) Population and agents’ attributes & \\
\hspace{2em} Create 3000 \textit{senior} agents & \\
\hspace{2em} Set location according to segregation & \\
\hspace{2em} Set seniors’ variables: age, income, educational status & \\
\hspace{2em} Set seniors \textit{hatch} a junior according to birth rate & \\
\hspace{2em} Set students’ variables: age, endowment, GPA & \\
\hspace{2em} Let students move 3 steps away from parents & \\
\hspace{1em} (b) Environment & \\
\hspace{2em} Set social reach 4.5 to identify neighbors & \\
\hspace{2em} Set importance \(\nu\) for agents and parents & \\
2. \textbf{Model dynamics} & \\
\hspace{2em} Iterate the model 100 \(t\). For each \(t\): & \\
\hspace{1em} (a) Students aged 17 take high-school exam & \\
\hspace{1em} (b) Decide whether to live at home or outliving during studies & \\
\hspace{1em} (c) Calculate budget (e.g. income, grant(s), endowment and loan) & \\
\hspace{2em} Interact with neighbors and compute: & \\
\hspace{2em} Expectations on future income & \\
\hspace{2em} Expectations on future consumption & \\
\hspace{2em} Peer Influence & \\
\hspace{2em} Student's Disposition & \\
\hspace{2em} Openness & \\
\hspace{2em} Distance & \\
\hspace{2em} Take a decision: & \\
\hspace{2em} Enroll and complete according to \(Pr_{\text{complete}}\) & \\
\hspace{2em} Change generational status into senior & \\
\hspace{2em} Obtain a educated/uneducated income & \\
\hspace{2em} Set location according to segregation & \\
\hspace{1em} (d) Seniors \textit{hatch} and die according to birth rate and carrying capacity & \\
\hspace{1em} (e) Set-up new juniors as in the initialization procedure & \\
\hspace{1em} (f) Update agents’ age by adding 1 year & \\
\hspace{1em} (g) Compute and update output variables & \\
3. \textbf{Monte Carlo experiment} & \\
\hspace{1em} (a) Define output variables of which results must be collected & \\
\hspace{1em} (b) Run 100 experiments using the Behavior Space & \\
\bottomrule
\end{tabular}
\end{adjustbox}
\end{table}

\section{Results}\label{sec:results}

\begin{figure}[!htbp]
  \centering
  \includegraphics[width=\linewidth]{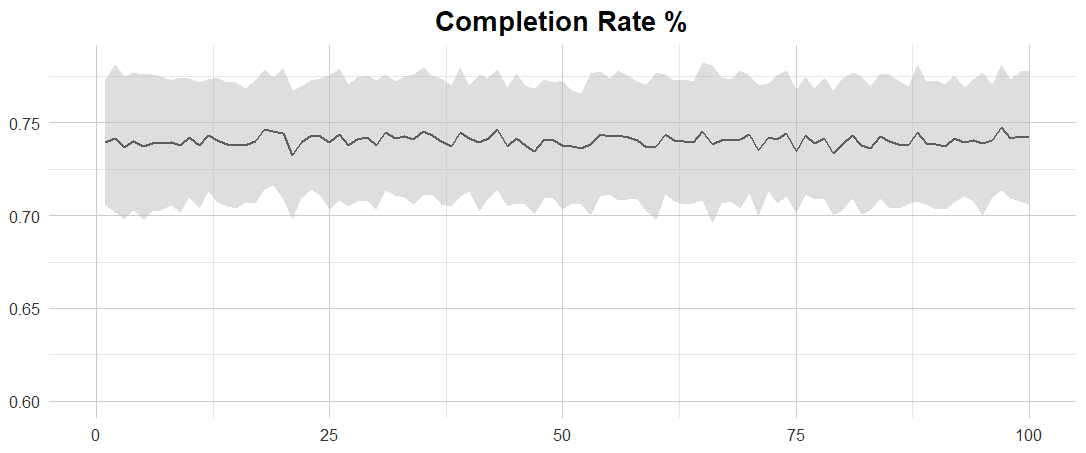}
  \caption{The average completion rate (as a percentage) over 100 Monte Carlo simulations is represented by a continuous line, with one standard deviation shown in grey, plotted across the simulated time span.}
  \label{fig:completion_rate}
\end{figure}

\begin{figure}[!htbp]
    \centering
    \includegraphics[width=\linewidth]{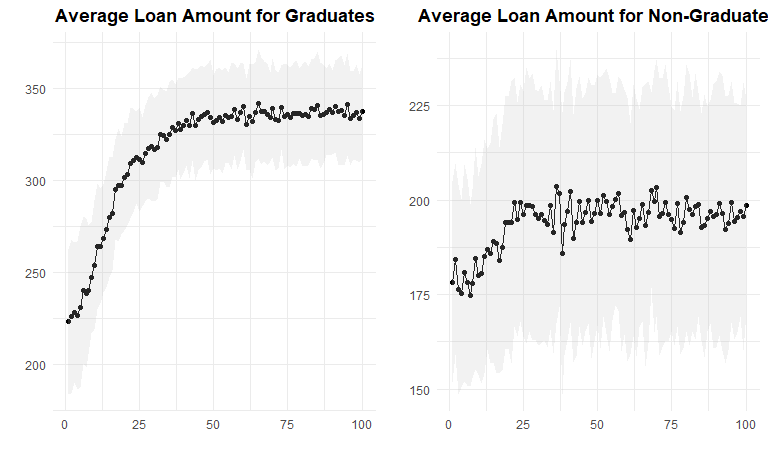}
    \caption{Average Loan Amount for Graduates and Non-Graduates (benchmark model) over 100 Monte Carlo simulations.}
    \label{fig:average_loan}
\end{figure}

The completion rate is the main indicator of interest, calculated as the percentage of cohort students who decided to enrol and eventually completed their higher education degree, and is shown in Figure \ref{fig:completion_rate}. Another indicator included is the share of students completing higher education coming from an educated or uneducated family, highlighting the importance of parental education. Lastly, the model presents the average amount loaned by students from uneducated and educated families (see Figure \ref{fig:average_loan}). 

Firstly, the  obtained completion rate from the simulation model averages out to 74.0 per cent, reasonably close to the 75.1 per cent that results from CBS data \citep{CBS2023}. Secondly, the simulation demonstrates a slightly higher completion rate for students enrolled as first-generation students - those whose parents did not pursue higher education - at 74.20 per cent compared to 73.9. At first glance, the second result might appear counter-intuitive since first-generational students receive less parental guidance and financial assistance in their educational journey. 

\cite{monitor2022} indeed shows a higher drop-out rate in the first year of applied and scientific university for first-generational students (16 to 13\% and 6 to 4\%). This evidence complements international research highlighting lack of guidance, financial and family responsibilities as main barriers for the lower retention rate \citep{engle2007postsecondary, falcon2015breaking}.

However, one should remember that the students in the simulation already made it to the final high school exam at the pre-university level from a disadvantaged position. Therefore, the students who lack the personal traits to compensate for a disadvantaged background might already be filtered out.

Thirdly, the simulation shows that first-generational students loan significantly less than their peers (197 versus 331 euros). In the model, work income is determined exogenously and the decision to live independently is based on risk preference. Thus, leaving the additional grant and parental financial support as the only two variables that could have caused this difference. Since parental financial support slowly increases across incomes, except for the distribution's tails (see Table \ref{table1}), the displayed difference is attributable to the additional grant. To verify this, Section \ref{sec:scenario} examines a setting in which additional grants are non-existent.  

The Dutch education monitor observes similar differences, with students from educated families gathering most of their income from parental contributions and loans, while first-generational students report larger labour income and receive higher additional grants \citep{monitor2022}. Note that undergraduates from bottom incomes are an exemption and loan significantly more than average low-income peers. These students often receive negligible parental contributions and lack financial literacy from home. 

Lastly, comparing the loan amounts obtained by the simulation with real-world data (average amount of 575 euros)  is difficult \citep{OCW2024lenen}. These include students not eligible for the basic grant reinstalled in September 2023. To verify the results of the NetLogo model, Section \ref{sec:scenario} evaluates a scenario that excludes the basic grant. These measures will be compared to loan data of 2022.

\section{Scenario Analysis}\label{sec:scenario}

Following from the discussion in the results section, this part examines three alternative scenarios:

\begin{enumerate}
    \item The first scenario evaluates the consequences of abolishing the additional grant for students from lower-income households. In particular, this experiment should verify that this policy is the major explanation for the difference between the average loan amount of graduates and non-graduates. 
    \item In the second scenario, the basic grant is abolished for all students, irrespective of their living situation. By following this approach, it is possible to observe the impact of a basic ground on average loan amounts. This is presented by advocates as the most detrimental reason to reinstall it, as student debts have a significant impact on student mental well-being and prospects \citep{van2022encouraging}. Additionally, the scenario allows us to compare the real world's and simulation's data on average loan amounts.
    \item Lastly, we introduce an abstract framework in which the relative wage for educated to uneducated jobs is set to one. Practically, this implies that prospective students do not observe any wage premium. The reason for evaluating the scenario is the large variability between wages within the educated and uneducated job market. Consequently, the possibility exists that prospective students interested in construction obtain a completely different reference point than those who observe other uneducated fields as alternatives for an educated living. To examine the impact of varying reference points, the extreme case of a relative wage equal to one is inspected.
\end{enumerate}

\begin{table}[h!]
\centering
\caption{Average values and corresponding standard deviations across 100 simulations, calculated over the time span 01--100, for the baseline model and the three alternative scenarios tested.}
\label{tab:3}
\scriptsize 
\begin{tabularx}{\textwidth}{lcccccccc}
\toprule
 & \multicolumn{2}{c}{\textbf{Baseline model}} & \multicolumn{2}{c}{\textbf{Scenario 1}} & \multicolumn{2}{c}{\textbf{Scenario 2}} & \multicolumn{2}{c}{\textbf{Scenario 3}} \\ \cmidrule(lr){2-3} \cmidrule(lr){4-5} \cmidrule(lr){6-7} \cmidrule(lr){8-9}
\textbf{} & \textbf{Mean} & \textbf{sd} & \textbf{Mean} & \textbf{sd} & \textbf{Mean} & \textbf{sd} & \textbf{Mean} & \textbf{sd} \\ \midrule
\textbf{Completion rate \%} & 74.0** & 3.38 & 73.8** & 3.38 & 72.4** & 3.38 & 54.0** & 3.80 \\ 
\textbf{Completion (edu.) \%} & 73.9** & 4.44 & 73.8 & 4.40 & 72.0** & 4.49 & 54.0* & 5.25 \\ 
\textbf{Completion (unedu.) \%} & 74.2** & 5.41 & 73.8 & 5.43 & 73.1** & 5.40 & 54.1* & 5.62 \\ 
\textbf{Loan amount (edu.) } & 317** & 43.02 & 402** & 48.84 & 480** & 53.25 & 320** & 47.16 \\ 
\textbf{Loan amount (unedu.) } & 194** & 32.93 & 398** & 46.13 & 356** & 48.74 & 210** & 37.03 \\ 
\bottomrule
\end{tabularx}
\vspace{0.2cm} 
\footnotesize
\textit{Notes:} ** significance at the 1\% level; * significance at the 5\% level.\\
Welch Two-Sample t-test conducted on enrol rate base versus one of the scenarios. Other t-tests results displayed are conducted within columns (e.g. enrolling (edu.) versus enrolling (unedu.) and loan amount (edu.) versus loan amount (unedu.))
\end{table}

As expected, abolishing the additional grant policy significantly increases the loan amounts of first-generational students (194 vs 398 euros). Although first-generation students are more reluctant to borrow than students from educated families, they are given no alternative as part-time labour income cannot fully cover living expenses. Concerning students with educated parents, Table \ref{tab:3} demonstrates an 85 euros increase in loan amount compared to the baseline model. This rise results from awarding partial additional grants to students from middle-income families \citep{duo2023}. Overall, eliminating the additional grant policy results in a significant convergence in loan amounts among students from different backgrounds.

While Scenario 1 had minimal impact on the completion rate, abolishing the basic grant shows a notable effect. The average completion rate drops by 1.6 percentage points, with students from educated families experiencing an above-average decrease of 1.9 percentage points. Relative to the baseline model, loan amounts rose by 51.4 and 83.5 per cent for students from educated and uneducated backgrounds respectively. From these outputs, it is observable that the installment of the basic grant did enhance the relative standing of first-generation students to their peers. The numbers shown in Table \ref{tab:3} align closely with those reported by the Ministry of Education (average of 435.11 euros) \citep{OCW2024lenen, CBS2020lenenwerken}.

Finally, neutralising the observed wage premium has a similar impact on the completion rate, regardless of the student's background. The completion rate decreased drastically by 27 per cent, which is a result in line with expectations. Contrary, the over 8 per cent increase in the average amount loaned by first-generation students is more notable. This suggests that those who switch earn relatively more from work income (or additional grant support) in the baseline model compared to other first-generation students. This indicates a greater emphasis on their expected consumption in comparison to their peers. 

The relevance of this abstract scenario is demonstrated by the developments in construction employment in recent years. As highlighted before, the construction sector offers wages above the level of several educated jobs, especially for those self-employed \citep{monitor2022}. This makes the sector an attractive alternative for prospective students who doubt college enrolment. In practice, employment in construction surged to almost 350.000, representing an increase of over 10 per cent from 2019 \citep{UWV2024}. Moreover, the number of self-employed builders rose by nearly 30 per cent in the same period \citep{monitor2022}. Surely, most of these jobs are filled by individuals who switch from different occupations and would otherwise not pursue a college education. However, this trend represents a general increased interest in construction work, which most likely also applies to prospective students. 

\section{Sensitivity Analysis}\label{sec:sensitivity}

\begin{figure}[H]
    \centering
    \includegraphics[width=\textwidth]{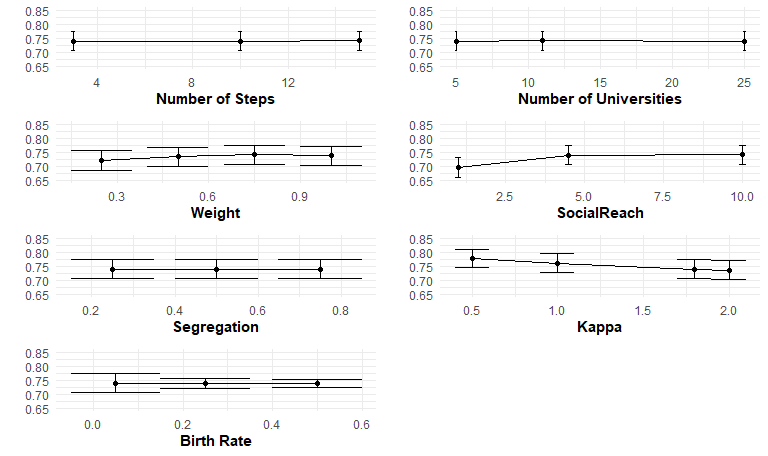}
    \caption{Sensitivity tests on the baseline model}
    \label{fig:sensitivity_tests}
\end{figure}

Several exogenous-determined parameters might affect the five variables included in the model. In this section, the degree of impact of these parameters is evaluated. The method used to perform the review is the one-at-a-time sensitivity analysis (OAT), which displays the model's response to changes in individual parameters. The output for all parameters is given in Figure \ref{fig:sensitivity_tests} and discussed per parameter in the coming paragraphs. 

The first parameter assessed is the \textit{number of steps} a prospective student takes away from their parent upon creation. This ensures that students' social connections partially overlap that of their parents, but are not entirely similar. Regarding the baseline model, this value is initially set to 3 while the parameter sweep includes 10 and 15. The output displays a minimal increase in the completion rate. The increase might result from students coming from practically educated backgrounds, whose parents are less likely to have significant relationships with successful, highly educated peers \citep{stevens2011cohort}. 

Varying the \textit{number of universities} from 5 to 11 to 25 produces the smallest difference in the completion rate. Increasing it from 5 to 11 lowers the value of the centrality fraction, which has a slight positive impact on the variable of interest. The following rise to 25 has a greater effect on maximum distance than average distance, leading to a rebound in the completion rate. 

Contrarily, shifting the \textit{weight} creates a notable variance in the completion rate. The parameter sweep includes the values 0.25, 0.5, 0.75 and 1, where the last one implies that the study decision is solely affected by the observed wage premium. As observable, the completion rate line follows a concave trend, indicating that increasing the weight has little to no effect after it reaches 0.75. The positive effect of weight on the completion rates shows that the combined impact of peers' influence, student's disposition, personality and centrality is less positive than that of wage premium alone. 

The \textit{social reach} parameter is evaluated for values equal to 1, 4 and 10. This parameter also exhibits a positive concave relationship with the completion rate. This might imply that increasing one's social reach produces a more representative image of wage differences between those practically and highly skilled. 

Furthermore, sweeping the \textit{segregation} value produces no observable changes to the completion rate. In specific, the values tested are 0.25, 0.5 and 0.75. It should be noted that all parameters are tested while holding all other factors constant at their baseline level. Therefore, the possibility exists that the segregation value interacting with another factor does demonstrate noteworthy differences. For example, a relatively low social reach value could increase the influence of the segregation parameter.

The $\kappa$ parameter reflects ability differences between high-achievers and others, where a larger $\kappa$ represents a greater disparity in returns for studying between the two groups. This implies that (below) average students are less likely to successfully attend college for increasing values of $\kappa$. Resultingly, the output for this parameter displays a linearly downward-sloping line. The value attributed in the baseline model equals 1.8 with additional tests conducted for the values 0.5, 1 and 2.   

Lastly, varying the \textit{birth rate} values does not demonstrate any impact on the average completion rate. The values included are 0.25, 0.5, 0.75. However, larger birth rates result in less variance in the dependent value outcomes. This is expected since a higher birth rate implies a higher sample of prospective students for each tick. 

Overall, sweeping one of the parameters while holding all else constant does not lead to large or unexpected changes in the average completion rate. The sensitivity analysis verifies that these parameters function as intended. For instance, the $k$ factor accurately shows a negative relationship with completion rates and insufficient social reach is disadvantageous for individuals from uneducated backgrounds. 

\section{Labour Market Dynamics}\label{sec:labor}

To this point, the paper has concentrated on identifying the factors contributing to the relatively high completion rate of tertiary education in the Netherlands. As an extension, this section will examine current labour market trends to investigate whether such a high completion rate is both desirable and sustainable. 

To begin with, it is important to note that the Dutch are not an exception to the typical age distribution seen in Western societies. Currently, there are 3 people of working age for 1 elderly person (65+) \citep{cbs2023elderly}. The ratio sharply decreased from 1 to 5 at the start of the century and is expected to decline further in the coming decades. This trend causes a tightening in the labour market for unskilled and skilled labour. 

Secondly, the transition from low-skilled labour to high-skilled labour due to technological advancements is projected to amplify in the coming years. Wage levels are currently not predominated by automation, but advancements in AI could alter this situation rapidly. Multiple studies indicate that automation leads to downward pressure on wages for the least educated, whereas highly educated jobs appear more robust \citep{acemoglu2020robots, borjas2019immigrants, xie2021ai, frey2017future}. 

For instance, \cite{acemoglu2020robots} focus on the influence of industrial robots on wage dynamics. Robots positively impact wages by increasing productivity and create a negative effect through the direct displacement of workers. The research captures the adverse effects by observing how employment within each occupation changes relative to the overall working population. The European data indicates that the negative impact is primarily concentrated in the manufacturing sectors. The most heavily robotised industries include automotive, chemicals and metals during the period 1990 to 2007. Contrary, numerous other low-skilled sectors such as agriculture, construction and services are barely impacted. 

In a related study, \cite{frey2017future} assess the impact of computerisation on job stability in the US. First, the research creates an index displaying the probability of computerisation based on the level of social intelligence, creative intelligence and coordinated movements needed. These variables are then used to determine the level of vulnerability to computerisation per occupation. The authors conclude that 47 per cent of US employment is in the high-risk category over some unspecified number of years. In comparison to \cite{acemoglu2020robots}, this study also stresses the impact on administration, transportation and services workers.  

Moreover, \cite{borjas2019immigrants} analysed data on the growth of robots in the US from 1996 to 2016. The increase in number of robots per worker did not have a significant impact on wage dynamics for American workers. However, the study emphasises that the observed exponential growth of robots can potentially disrupt the job market. The introduction of AI accelerates this process.  

In contrast, \cite{kalyani2023stakeholders} observe significant layoffs in the American tech industry, primarily affecting highly skilled workers. Nonetheless, the book's chapter recommends that companies retain these individuals and seek suitable alternative roles, highlighting their crucial role in driving the company's growth. In general, the outlined studies indicate that practical skilled labour bears the brunt of automation, although the extent of the impact varies depending on occupation and ongoing advancements in AI. 

Thirdly, the Dutch Planning Agency projects a decrease in the number of low-skilled jobs but an even faster decrease in the supply of unskilled labour \citep{simic2015labour}. This is a critical issue as practical skills are needed in manufacturing, healthcare and to complete the energy transition. A labour shortage in these sectors applies additional pressure on the housing and energy crises. With this in mind, the Dutch Minister of Education urged pre-university graduates to consider secondary vocational education. 

Furthermore, the report foresees an increase in skilled labour supply, with an even more rapid surge in skilled work vacancies. Similar results were obtained by the Dutch Ministry of Foreign Affairs, which investigated the option of resolving labour market shortages in ICT and higher-technical jobs by promoting migration from African and Middle Eastern students \citep{oomes2019dutch}. 

In conclusion, this discussion demonstrated the complexity of the issue. Regarding desirability, the main matter is the general scarcity of labour. While many practical positions are more likely to be automated, labour demand in vital sectors such as healthcare and construction is increasing \citep{simic2015labour}. Simultaneously, the number of skilled workers must grow to maintain a top-tier position in technology-related industries. To meet this demand, even higher completion rates are desired, though this appears unrealistic. Therefore, the most intuitive solution to address both shortages is labour migration \citep{simic2015labour}. However, this approach brings about other societal issues outside the scope of this paper. 

Moreover, sustainability mostly depends on developments in scarcity as well. For instance, if relatively high wages in construction persist, it is very plausible that more and more students will decide on a practical/vocational education instead of a theoretical one. 

\section{Conclusion}\label{sec:conclusion}

The purpose of this study was to identify the primary factors contributing to the high completion rate in the Dutch tertiary educational system. Firstly, agent-based modelling was selected as the preferred method to investigate the research question.  This approach offers a distinct advantage over other techniques due to its ability to simulate complex interactions among agents. In addition, the adaptability of the model's boundaries and exogenous parameters allows for convenient testing of policy changes. 

A literature review was then conducted to identify the possible variables of influence. The uniqueness of this research is the inclusion of psychological and sociological factors among economic variables. Eventually, wage premium was identified as the primary economic variable, while student's disposition, peers' influence, personality and location were considered important personal attributes. Given that Nordic countries emphasise job security and payment more than other European countries, a weight of 0.75 was assigned to the wage premium variable. 

Following a discussion of the model's exogenous parameters and dynamics, the results provided an average completion rate consistent with real-world observations. However, the model revealed a slightly higher completion rate among first-generation students than for those from educated backgrounds, thereby contrasting with actual retention rates. The significant difference in loan amount between the two categories is close to that observed in reality. Students from educated backgrounds may feel more secure taking on loans, as their families typically have better financial stability. 

When testing policy alternatives, the difference in loan amount was erased when eliminating the additional grant. In this scenario, first-generation students were compelled to take out loans, as the basic grant combined with working was insufficient to meet financial needs. Furthermore, erasing observed wage premium had significant negative effects on completion rates, suggesting that prospective students who view construction as an alternative may be less inclined to pursue higher education. 

The discussion on labour market dynamics highlighted that shortages exist in both educated and practical jobs, making it unlikely that these issues can be resolved solely by altering the completion rate of tertiary education. Moreover, the development of these shortages influences the observed wage premium, which in turn plays a critical role in whether enrolling remains desirable for prospective students. 

The research also identifies promising areas for future research. In this study, completion rates from both applied and theoretical universities were combined. However, theoretical universities report significantly higher completion rates and lower dropout rates than applied universities \citep{CBS2023}. Exploring the reasons behind this disparity could improve policies aimed at reducing drop-out rate, yielding numerous benefits.

Currently, the drop-out rate included in the model is treated as an exogenous parameter. Treating the variable as endogenous would help verify whether the higher drop-out rate among first-generation students is indeed caused by a lack of guidance and financial resources. This adjustment would also enhance the model's overall accuracy, particularly in predicting completion rates.

Looking ahead, the recently installed government has detailed plans to discourage student migration into the Netherlands \citep{HOOfdlijnenakkoord}.
This policy initiative is mainly driven by the strain international students place on housing, as the market is already completely out of balance. Furthermore, only 24 per cent of international graduates remain in the country five years after completing their studies \citep{nuffic2023}

As imaginable, this might have significant implications for educational attainment and quality. For instance, several faculties of border universities (e.g. Maastricht, Twente) rely on international students to sustain certain programs and specialisations. Without these students, these programs and the expertise of affiliated scholars might be lost. Consequently, regulating student migration could negatively affect the quality and availability of education for domestic students.

\newpage
\appendix
\renewcommand{\thetable}{A\arabic{table}}
\setcounter{table}{0}

\section*{Appendix A}
\begin{table}[H]
    \centering
    \caption{Average Gross Salary Per Occupation}
    \label{table1}
%    \adjustbox{max width=\textwidth}{
    \begin{tabular}{ccc} 
    \toprule
        \textbf{Occupation} & \textbf{Average Gross Monthly Wage} & \textbf{Level} \\
        \midrule
        Climate Change Officer &  € 3,338.00  & Educated \\
        Teacher &  € 3,359.00  & Educated \\ 
        Programmer &  € 3,650.00  & Educated \\ 
        Human Resource &  € 3,777.00  & Educated \\ 
        Software Developer &  € 3,815.00  & Educated \\
        Researcher (WO) &  € 3,858.00  & Educated \\ 
        Accountancy \& Finance &  € 3,863.00  & Educated \\ 
        Basic Doctors  &  € 3,887.00  & Educated \\ 
        Marketing &  € 3,915.00  & Educated \\ 
        Architect &  € 4,251.00  & Educated \\ 
        Management &  € 4,706.00  & Educated \\ 
        Occupational Physician &  € 5,000.00  & Educated \\ 
        Psychologist &  € 5,563.00  & Educated \\ 
        Lawyer &  € 6,500.00  & Educated \\ 
        General Practitioner &  € 8,150.00  & Educated \\ 
        Facilitair &  € 2,326.00  & Mixed \\ 
        Administration &  € 2,771.00  & Mixed \\ 
        Designer &  € 2,844.00  & Mixed \\ 
        Sales &  € 3,081.00  & Mixed \\ 
        Media &  € 3,220.00  & Mixed \\ 
        Nurse &  € 3,406.00  & Mixed \\
        Social Worker &  € 3,909.00  & Mixed \\ 
        Pedagogical &  € 2,238.00  & Practically \\ 
        Beauty Care &  € 2,332.00  & Practically \\ 
        Customer Service &  € 2,489.00  & Practically \\ 
        Production &  € 2,515.00  & Practically \\ 
        Caretaker &  € 2,555.00  & Practically \\ 
        Doctor's Assistant &  € 2,615.00  & Practically \\ 
        Transportation &  € 2,855.00  & Practically \\ 
        Carpenter &  € 2,900.00  & Practically \\ 
        Painter &  € 3,000.00  & Practically \\ 
        Electrician &  € 3,050.00  & Practically \\ 
        Plumber &  € 3,100.00  & Practically \\ 
        Security &  € 3,116.00  & Practically \\ 
        Contractor &  € 3,350.00  & Practically \\ 
        Executor (Construction) &  € 3,900.00  & Practically \\ 
        \bottomrule
    \end{tabular}
%    }
    \smallskip
    \footnotesize Sources: \citep{NationaleBeroepenGids2023}

\end{table}
\begin{table}[htbp]
    \centering
    \caption{Parental Endowment}
    \label{table2}
 %   \adjustbox{max width=\textwidth}{
    \begin{tabular}{cccc}
        \toprule
        \textbf{Parental Income (avg.)} & \textbf{Income Distribution \%} & \textbf{ChildEdu \%} & \textbf{Absolute Spending} \\
        \midrule
        € 13800 & 0 - 10 & 16.3 & € 2249.40\\
        € 24300 & 10 - 20 &  8.2 & € 1992.60\\
        € 32000 & 20 - 30  & 6.9 & € 2208.00\\
        € 41000 & 30 - 40 &  5.1 & € 2091.00\\
        € 52000  & 40 - 50   & 5.2 & € 2704.00\\
        € 65200 & 50 - 60  & 5.4 & € 3520.8\\
        € 80900  & 60 - 70 & 5.0 & € 4045.00\\
        € 99900 & 70 - 80 & 5.3 & € 5294.70\\
        € 127100 &  80 - 90 & 5.8 & € 7371.80\\
        € 229200 & 90 - 100 &  5.6 & € 12835.20\\
        \bottomrule
    \end{tabular}
%    }
    \smallskip
    \footnotesize Sources: \citep{CBS2022InkomenHuishoudens} \& \citep{kornrich2013investing}
\end{table}

\begin{table}[htbp]
    \centering
    \caption{Monthly Living Expenses Non-Resident Students}
    \label{table3}
 %   \adjustbox{max width=\textwidth}{
    \begin{tabular}{ccc}
        \toprule
        \textbf{Expenses} & \textbf{Amount (€)} \\
        \midrule
        Rent  & 494\\
        Groceries  & 201\\
        Study Materials & 57 \\
        Tuition Fee  & 211 \\
        Leisure & 144\\
        Clothing & 62\\
        Transport  & 84\\
        Telephone & 22\\
        Vacation & 169\\
        \textbf{Total} & \textbf{1258}\\
        \bottomrule
    \end{tabular}
%    }
    \smallskip
    \footnotesize Sources: \citep{Nibud2021} \& \citep{CBS2024Inflatie}
\end{table}

\bibliography{ref}

\end{document}